\documentclass[preprint,pre,showpacs,superscriptaddress]{revtex4}
\usepackage{graphicx}
\usepackage{dcolumn}
\usepackage{bm}

\begin{document}
\title{On the exact solution of the mixed-spin Ising chain with 
axial and rhombic zero-field splitting parameters}
\author{M. DAN\v{C}O} 
\author{J. STRE\v{C}KA} 
\affiliation{Department of Theoretical Physics and Astrophysics, Faculty of Science, \\ 
P. J. \v{S}af\'{a}rik University, Park Angelinum 9, 040 01 Ko\v{s}ice, Slovak Republic} 
      
\begin{abstract}
Ground-state phase diagram of the mixed spin-$1/2$ and spin-$1$ Ising chain with axial and rhombic zero-field splitting parameters is exactly calculated within the framework of the transfer-matrix method. It is shown that the rhombic zero-field splitting parameter prefers the magnetically ordered phase instead of the disordered phase.
\end{abstract}

\maketitle

\section{Introduction}
Exactly solved one-dimensional quantum spin models traditionally belong to the most fascinating research areas as they provide valuable insight into otherwise hardly understandable aspects of cooperative and quantum phenomena \cite{matt93}. In this work, we will exactly treat the mixed
spin-1/2 and spin-1 Ising chain with axial and rhombic zero-field splitting parameters.

\section{Model and its exact solution}
Consider the Ising model for a chain consisting of the alternating spin-$1/2$ and spin-$1$ atoms, 
which accounts also for axial and rhombic zero-field splitting parameters. The total Hamiltonian of this spin system can be written as a sum of two terms $\hat{\cal H} = \hat{\cal H}_{ex} + \hat{\cal H}_{zfs}^{(1)}$. The former term accounts for the Ising-type exchange interaction between the nearest-neighbor spins  
\begin{eqnarray}
\hat{\cal H}_{ex} = -J \sum_{k=1}^{N} \hat{S}_{k}^z (\hat{\sigma}_{k}^z + \hat{\sigma}_{k+1}^z),
\label{1}
\end{eqnarray}
and the latter term accounts for the axial ($D$) and rhombic ($E$) zero-field splitting (ZFS) parameters acting on the spin-$1$ atoms only
\begin{eqnarray}
\hat{\cal H}_{zfs}^{(1)} = -D \sum_{k=1}^{N} (\hat{S}_{k}^z)^2 
                           - E \sum_{k=1}^{N} [(\hat{S}_{k}^x)^2 - (\hat{S}_{k}^y)^2].
\label{2}
\end{eqnarray}
Above, $\hat{\sigma}_{k}^z$ and $\hat{S}_{k}^\alpha (\alpha=x,y,z)$ denote standard spatial components of the spin-$1/2$ and spin-$1$ operators, respectively, $N$ denotes a total number of spin-$1/2$ (spin-$1$) atoms and the periodic boundary condition $\sigma_{N+1}\equiv \sigma_{1}$ is imposed 
for further convenience. It is worthwhile to remark that there is one-to-one correspondence between 
the Hamiltonian $\hat{\cal H}_{zfs}^{(1)}$ given by Eq.~(\ref{2}) and the Hamiltonian with three different parameters $D^x$, $D^y$ and $D^z$
\begin{eqnarray}
\hat{\cal H}_{zfs}^{(2)} = - D^x \sum_{k=1}^{N} (\hat{S}_{k}^x)^2 
                           - D^y \sum_{k=1}^{N} (\hat{S}_{k}^y)^2 
                           - D^z \sum_{k=1}^{N} (\hat{S}_{k}^z)^2.
\label{3}
\end{eqnarray}
The equivalence between $\hat{\cal H}_{zfs}^{(1)}$ and $\hat{\cal H}_{zfs}^{(2)}$ can easily be verified by establishing a rigorous mapping correspondence between the relevant interaction terms appearing in the Hamiltonians (\ref{2}) and (\ref{3}). The total angular momentum of the spin-$1$ atoms is integral of motion $\hat{\bf{S}}_{k}^2 = (\hat{S}_{k}^x)^2 + (\hat{S}_{k}^y)^2 + (\hat{S}_{k}^z)^2 = 2$ and hence, one of three parameters $D^x$, $D^y$ and $D^z$ must depend on the other two. Consequently, the Hamiltonians $\hat{\cal H}_{zfs}^{(1)}$ and $\hat{\cal H}_{zfs}^{(2)}$ differ one from the other just by some constant factor $\hat{\cal H}_{zfs}^{(1)} = \hat{\cal H}_{zfs}^{(2)} + C$, whereas the relevant interaction terms $C$, $D$ and $E$ are connected to the ones 
$D^x$, $D^y$ and $D^z$ through the mapping relations
\begin{eqnarray}
C = D^x + D^y, \, D = D^z - \frac{D^x + D^y}{2}, \, E = \frac{D^x - D^y}{2}.
\label{4}
\end{eqnarray}
The model under investigation thus turns out to be equivalent to the one recently studied 
by Wu \textit{et al.} \cite{wu08} using the approach based on Jordan-Wigner transformation. 

Here, the investigated model system will be exactly treated within the framework of transfer-matrix method \cite{kram41}. First, it is useful to rewrite the total Hamiltonian as a sum of site Hamiltonians $\hat{\cal H} = \sum_k \hat{\cal H}_{k}$, whereas each site Hamiltonian $\hat{\cal H}_{k}$ 
involves all the interaction terms associated with the spin-$1$ atom from the $k$th lattice site
\begin{eqnarray}
\hat{\cal H}_{k} = - J\hat{S}_{k}^z(\hat{\sigma}_{k}^z + \hat{\sigma}_{k+1}^z) 
                   - D(\hat{S}_{k}^z)^2 - E[(\hat{S}_{k}^x)^2 \!-\! (\hat{S}_{k}^y)^2].
\label{5}
\end{eqnarray}
Due to a validity of commutation relation between different site Hamiltonians, 
the partition function can be partially factorized into the product 
\begin{eqnarray}
{\cal Z} = \sum_{\{\sigma_{k}\}}\prod_{k=1}^{N} \mbox{Tr}_{S_{k}}\exp (-\beta \hat{\cal H}_{k}),
\label{6} 
\end{eqnarray}
where $\beta = 1/(k_{\rm B} T)$, $k_{\rm B}$ is Boltzmann's constant, $T$ is the absolute temperature, $\mbox{Tr}_{S_{k}}$ means a trace over spin degrees of freedom of the $k$th spin-$1$ atom and $\sum_{\{\sigma_{k}\}}$ denotes a summation over all possible configurations of the spin-$1/2$ atoms. After tracing out spin degrees of freedom of the spin-$1$ atom, the relevant expression on r.h.s of Eq.~(\ref{6}) will depend just on its two nearest-neighbor spins $\sigma_{k}$ and $\sigma_{k+1}$. Moreover, this expression can be subsequently used in order to define 
the transfer matrix
\begin{eqnarray}
T \!\!\!\!\!\!&&\!\!\!\!\!\! (\sigma_{k}, \sigma_{k+1}) 
= \mbox{Tr}_{S_{k}}\exp (-\beta \hat{\cal H}_{k}) \nonumber \\ \!\!\!&=&\!\!\! 
1 + 2 \exp(\beta D) \cosh \left(\beta \sqrt{J^2(\sigma_{k}^z + \sigma_{k+1}^z)^2 
  + E^2}\right)\!.
\label{7}
\end{eqnarray}
The rest of our exact calculations can be accomplished using the standard procedure developed within the transfer-matrix approach \cite{kram41}. This rigorous technique allows one to express the partition function in terms of respective eigenvalues of the transfer matrix
\begin{eqnarray}
{\cal Z} = \sum_{\{\sigma_{k}\}} \prod_{k=1}^{N} T(\sigma_{k}^z, \sigma_{k+1}^z) 
= \mbox{Tr} T^N = \lambda_{+}^N + \lambda_{-}^N.
\label{7b}
\end{eqnarray}
In the thermodynamic limit $N \to \infty$, the free energy per unit cell can be expressed solely 
in terms of the largest eigenvalue of the transfer matrix
\begin{eqnarray}
f = - k_{\rm B} T \lim_{N \to \infty} \frac{1}{N} \ln {\cal Z} = - k_{\rm B}T \ln (T_{11} + T_{12}),
\label{8}
\end{eqnarray}
where $T_{11} = T (\pm 1/2, \pm1/2)$ and $T_{12} = T(\pm 1/2, \mp1/2)$ were used to denote two different matrix elements of the transfer matrix defined through Eq.~(\ref{7}).

\section{Results and discussion}

Now, let us take a closer look at the ground-state behavior of the investigated model system. For simplicity, our subsequent analysis will be restricted only to the particular case with 
the ferromagnetic interaction $J>0$, since the relevant change in sign of the parameter 
$J$ causes just a rather trivial reversal of all the spin-1/2 atoms.

In the zero temperature limit, the first-order phase transition line given by the condition $D = -\sqrt{J^2 + E^2}$ separates the ferromagnetically ordered phase (OP) from the disordered phase (DP). The relevant spin order appearing in the OP and DP can be unambiguously defined through 
the eigenvectors
\begin{eqnarray}
\left| OP \right \rangle &=& \otimes_{k} \left| 1/2 \right \rangle_{k} 
 \left[\cos \left(\frac{\varphi}{2}\right) \left|+1 \right \rangle_{k} 
     + \sin \left(\frac{\varphi}{2}\right) \left|-1 \right \rangle_{k} \right], \nonumber \\
\left| DP \right \rangle &=& \otimes_{k} \left| \pm 1/2 \right \rangle_{k} 
                                         \left| 0 \right \rangle_{k},
\nonumber
\end{eqnarray}
where the product runs over all lattice sites, the former (latter) ket vectors specify the state 
of the spin-$1/2$ (spin-$1$) atoms and the mixing angle $\varphi$ is given by $\varphi = \arctan (E/J)$. In the DP, all the spin-$1$ atoms tend toward their 'non-magnetic' spin state $\left| 0 \right \rangle$ on behalf of a sufficiently strong (negative) axial ZFS parameter and hence, each spin-$1/2$ atom may completely independently choose any of two available spin states $\left| \pm 1/2 \right \rangle$. However, the more striking spin order emerges in the OP, where the magnetic behavior of the spin-$1$ atoms is governed by a quantum entanglement of two magnetic spin states $\left| +1 \right\rangle$ and $\left| -1 \right\rangle$ and all the spin-$1/2$ atoms reside their "up" 
spin state $\left| 1/2 \right\rangle$. In this respect, the rhombic ZFS parameter gradually 
destroys a perfect ferromagnetic order between the spin-1/2 and spin-1 atoms, which appears 
in an absence of the rhombic term. 
\begin{figure}[thb]
\includegraphics[width=6cm]{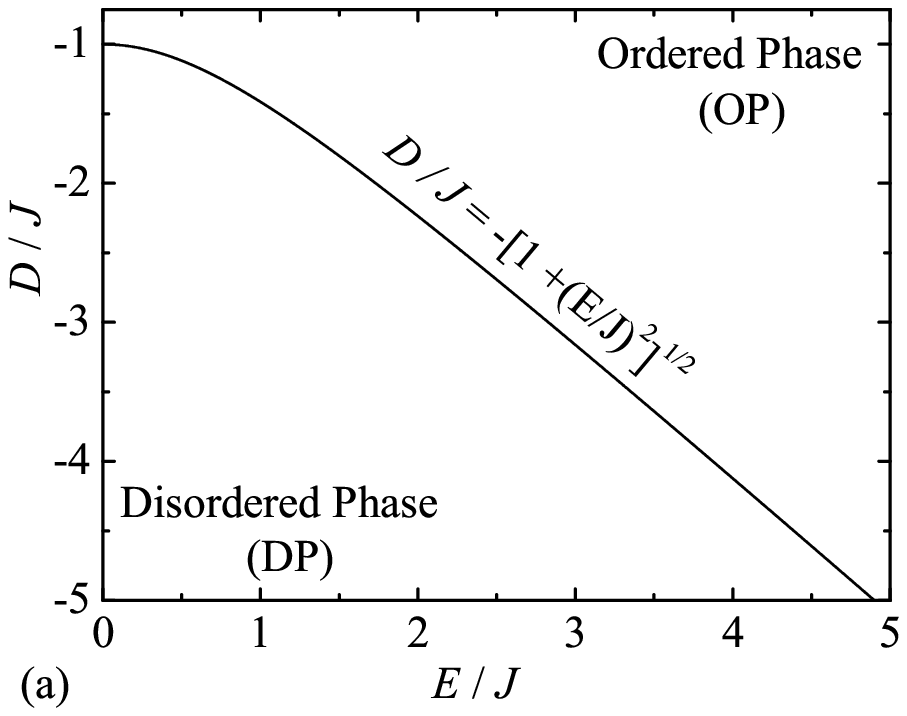}
\includegraphics[width=6cm]{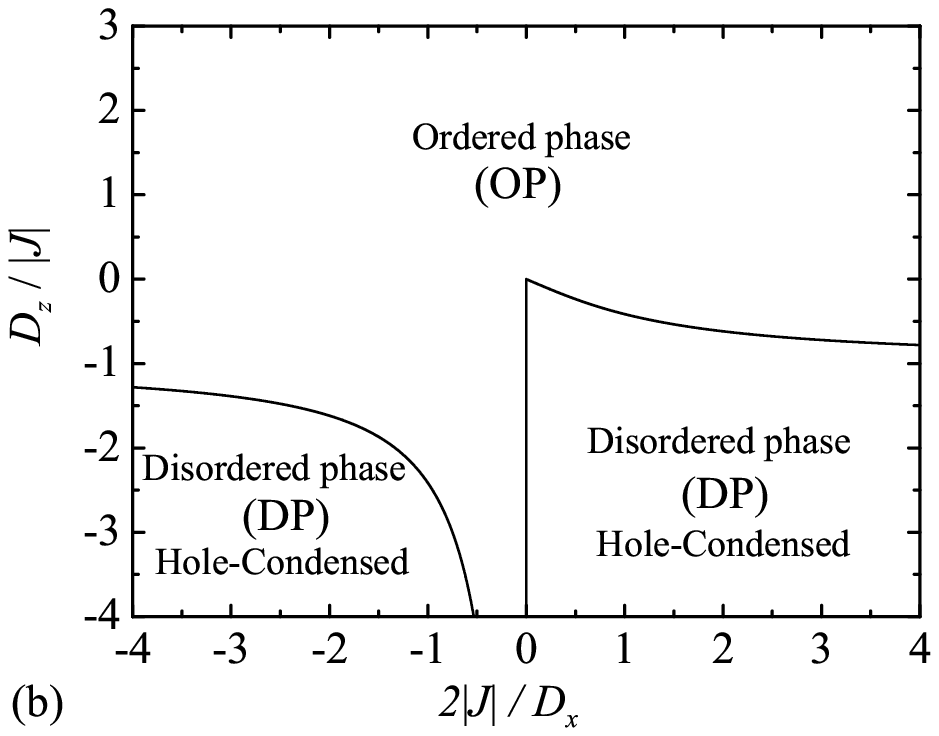}
\vspace{-0.7cm}
\caption{Ground-state phase diagram in two different planes: (a) $E/J - D/J$ plane; 
$(b)$ $2J/D_{x} - D_{z}/J$ plane for $D_y = 0$.}
\label{fig1}
\end{figure}

For better illustration, Fig.~\ref{fig1}(a) depicts the ground-state phase diagram in the $E/J-D/J$ plane. The most surprising finding stemming from Fig. \ref{fig1}(a) is that the phase boundary between OP and DP shifts toward more negative values of the axial ZFS parameter when increasing a strength 
of the rhombic ZFS parameter. Accordingly, it turns out that the quantum entanglement between 
the spin states $\left| +1 \right\rangle$ and $\left| -1 \right\rangle$, which is caused solely 
by the rhombic ZFS parameter, energetically stabilizes the OP before the DP. For comparison, 
Fig. \ref{fig1}(b) illustrates the  ground-state phase diagram in the $2J/D_{x}-D_{z}/J$ plane 
when using Eq. (\ref{3}) in order to define the ZFS Hamiltonian. Note that this phase diagram is 
in accord with the recent results of Wu \textit{et al}. \cite{wu08}, but this phase diagram is apparently less convenient for interpreting the phase boundary between OP and DP as the parameter $D_{x}$ changes according Eq.~(\ref{4}) both axial as well as rhombic ZFS parameters. 

In conclusion, it is worthy to notice that the rigorous procedure developed on the grounds of the transfer-matrix method can readily be adapted to treat the investigated model system even in a presence of non-zero external magnetic field, which will be examined in detail in our forthcoming work.

{\bf Acknowledgments}: This work was supported under the grant VEGA 1/0128/08.

\end{document}